\documentstyle[prb,aps,twocolumn,epsf,eqsecnum]{revtex}

\begin{document}

\draft

\twocolumn[\hsize\textwidth\columnwidth\hsize\csname @twocolumnfalse\endcsname

\title{Nucleation and Growth of the Superconducting Phase in 
the Presence of a Current}

\author{Andrew J. Dolgert,$^{1,}$\cite{emaild}
Thomas Blum,$^{1,}$\cite{emailt}
Alan T. Dorsey,$^{2,}$\cite{emaila} and 
Michael Fowler$^{1,}$\cite{emailm}}

\address{$^1$Department of Physics,  University of Virginia, McCormick 
Road, Charlottesville, Virginia 22901}

\address{$^2$Department of Physics, University of Florida, Gainesville, 
Florida 32611}

\date{\today}

\maketitle

\begin{abstract}
We study the localized stationary solutions of the one-dimensional
time-dependent Ginzburg-Landau equations in the presence of a current. 
These {\it threshold perturbations} separate undercritical 
perturbations which return to the normal phase from overcritical 
perturbations which lead to the superconducting phase. 
Careful numerical work in the small-current limit shows that
the amplitude of these solutions is exponentially small in the
current; we provide an approximate analysis which captures this
behavior.
As the current is increased toward the stall current $J^*$,
the width of these solutions diverges resulting in widely separated 
normal-superconducting interfaces. 
We map out numerically the dependence of $J^*$ on $u$ (a parameter
characterizing the material) and use asymptotic analysis to derive
the behaviors for large $u$ ($J^*\sim u^{-1/4}$) and small $u$
($J \rightarrow J_c$, the critical deparing current), which agree
with the numerical work in these regimes. 
For currents other than $J^*$ the interface moves, and in this case 
we study the interface velocity as a function of $u$ and $J$.  
We find that the velocities are bounded both as $J \rightarrow 0$ 
and as $J \rightarrow J_c$, contrary to previous claims. 
\end{abstract}


\vskip2pc]
 
\section{Introduction}

When a superconductor is placed in a magnetic field equal to its 
critical field $H_c$, the normal and superconducting phases can 
coexist in a state of equilibrium with the two phases separated 
by normal-superconducting (NS) interfaces.  
 The dynamics of such interfaces is important for various 
nonequilibrium phenomena. 
 For instance, if the applied magnetic field is quenched below $H_c$, 
these interfaces move through the sample, expelling the magnetic 
flux so as to establish the Meissner phase  
\cite{frahm91,liu91,dibartolo96,dorsey94,osborn94,chapman95a,%
pippard50,chapman95b,goldstein96}.
 Just as superconductivity can be destroyed by applying a magnetic
field exceeding $H_c$, it can also be destroyed by applying a current
exceeding the critical depairing current $J_c$.  
 Thus by analogy with the magnetic field case, one might expect a 
special value of the applied current $J^*<J_c$ at which the
superconducting and normal phases coexist, separated by a 
stationary NS interface \cite{likharev74}.  
 In contrast to the magnetic field induced NS interfaces, these 
current-induced NS interfaces are intrinsically nonequilibrium 
entities, and their structure depends upon the {\it dynamics} of 
the order parameter and magnetic field.
 The evolution and dynamics of these nonequilibrium interfaces is
the subject  of this paper. 

The current-induced NS interfaces arise in several contexts. 
 First, they are known to be important in understanding the dynamics
of the ``resistive state'' in superconducting wires and films (for a
review see \cite{ivlev84}), and in determining the global stability of
the normal and superconducting phases in the presence of a current
\cite{kramer77}. 
 Second, Aranson {\it et al.\/} \cite{aranson96} have recently studied
the nucleation of the normal phase in thin type-II superconducting
strips in the presence of both a magnetic field and a transport current.  
 They found that a sufficient current produced large normal droplets
containing multiple flux quanta.
 Without a current one finds stationary,  singly quantized vortices, 
with a larger amount of NS interface per flux quantum than 
a multiply quantized droplet.
 They conclude that the current produces an effective surface tension
for the NS interface which is positive, stabilizing the interface and
producing larger droplets with smaller surface area.  
 Motivated in part by its role in this phenomenon we wanted to 
re-examine the nonequilibrium stabilizing effects of current. 

Even when the superconducting phase is ostensibly the equilibrium
phase, a current makes the normal phase metastable, i.e., linearly
stable to infinitesimal superconducting perturbations. 
 A localized superconducting perturbation of finite amplitude, 
on the other hand, 
has one of two fates:  (1) its amplitude may ultimately 
shrink to zero restoring the normal phase (undercritical) or 
(2) it may grow eventually establishing the superconducting state 
(overcritical). 
 Separating these two possibilities are the {\it critical nuclei} 
or {\it threshold perturbations}, for present purposes stationary 
solutions of the time-dependent Ginzburg-Landau (TDGL) equations 
localized around the normal state. 
 As one raises the current, the amplitude of the threshold solution 
grows, implying that the normal phase becomes increasingly stable. 

At very low currents, the widths of the critical nuclei shrink as
the current is increased, but eventually this trend is reversed and
the width grows as the current is increased further. 
 In fact, as the current approaches a particular value, $J^*$ (the 
``stall current"\cite{aranson96}), the width diverges resulting in 
two well-separated, stationary NS interfaces.
 The interface solutions have been studied numerically by Likharev
\cite{likharev74}, who found that the interfaces were stationary at
$J^* \approx 0.335$ for $u=5.79$, where $u$ characterizes the material
and is $5.79$ for nonmagnetic impurities\cite{schmid66}. 
 They were also studied by Kramer and Baratoff \cite{kramer77}, who
found $J^* \approx 0.291$ for $u=12$ (corresponding to paramagnetic
impurities\cite{gorkov68}). 
 However, we know of no systematic study of the dependence of $J^*$ 
upon $u$. 
 In this work we remedy this situation by using a combination of 
numerical methods and analysis including matched asymptotic expansions
\cite{mae}. 
 We show that $J^* \sim u^{-1/4}$ for large $u$ in contrast to
a previous conjecture \cite{likharev74}, and we find how
$J^*$ approaches $J_c$ in the small-$u$ limit. 

At currents other than $J^*$, the interfaces move with a constant
velocity with the superconducting phase invading the normal phase
for $J<J^*$ and vice versa for $J^*<J<J_c$.      
 At currents close to $J^*$, the interface velocity is proportional
to $(J-J^*)$.
 Likharev \cite{likharev74} defined the constant of proportionality,
$\eta=(dc/dJ)^{-1}|_{J=J^*}$, where $c$ is the interface speed; he
found $\eta \approx 0.7$ for $u=5.79$. 
 In the extreme limits, $J \rightarrow 0$ and $J \rightarrow J_c$ 
Likharev predicted that the speed $c$ diverges.
 We find $c$ to be bounded in both cases and provide an analytic
expression for it as $J \rightarrow 0$.  

The results of this work are summarized in Table~\ref{table1}. 
 The rest of the paper is organized as follows. 
 After briefly reviewing the TDGL equations and the approximations 
used in this work (section \ref{TDGLsection}), we study the critical 
nuclei focusing on their size and shape in the limit $J \rightarrow 
0$ (section \ref{nuclei}). 
 We then move on to consider the stationary interface solutions; 
in particular we map out the dependence of the stall current $J^*$ 
on $u$ and supplement the numerical work with analysis of the 
$u \rightarrow \infty$ and $u \rightarrow 0$ limits (section 
\ref{stationary}). 
 Next, we examine moving interfaces first in the linear response 
regime and then in the limits $J \rightarrow 0$ and $J \rightarrow 
J_c$ (section \ref{moving}).   
 The appendix contains a calculation of the amplitudes of the
critical nuclei in the $J \rightarrow 0$ limit. 

\begin{table}
\caption{Summary of the primary results.} 
\label{table1}
\begin{tabular}{lll}
I. Critical nucleus & \\
\ \ \ \ \ Small-$J$ width & $W \sim (uJ)^{-1/2}$ & Sec. 
\ref{nuc-smallj} \\
\ \ \ \ \ Small-$J$ amplitude  & $\psi_0 \sim {\rm exp}\{-A/uJ \}$ & 
Sec.~\ref{nuc-smallj} \\ 
II. Stall current $J^*$            &     &   \\
\ \ \ \ \ Large-$u$ & $J^* = 0.584491 ~u^{-1/4}$ & 
Sec.~\ref{stat-largeu} \\
\ \ \ \ \ Small-$u$ & $J^* =  J_c
{\textstyle (1-u/8)^{1/2}\over \textstyle (1-u/24)^{3/2}}$ &
Sec.~\ref{stat-smallu} \\
III. Kinetic coeff. $\eta$ & & \\
\ \ \ \ \ Large-$u$ & $\eta  = 0.797 ~u^{3/4}$ & Sec.~\ref{moving} \\
\ \ \ \ \ Small-$u$ & $\eta \sim u^{3/2}$ & Sec.~\ref{moving} \\ 
IV. Interface speed & & \\
\ \ \ \ \ $J \rightarrow 0$ & $c \rightarrow 2/u$ &
Sec.~\ref{moving} \\
\ \ \ \ \ $J \rightarrow J_c$ & $c \sim u^{1/2}$ \ \ \ \ 
$(u \rightarrow 0)$  & Sec.~\ref{moving} \\
\ \ \ \ \  & $c \sim u^{-0.85}$ \ \ $(u \rightarrow \infty)$
& Sec.~\ref{moving}
\end{tabular}
\end{table}

\section{The TDGL equations}
\label{TDGLsection}

The starting point for our study is the set of TDGL equations for 
the order parameter $\psi$, the scalar potential $\Phi$, and the 
vector potential ${\bf A}$: 
\begin{eqnarray}
\hbar \gamma \left( \partial_t  
+ {i e^* \Phi \over \hbar} \right) \psi &=& 
\frac{\hbar^2}{2m}\left({\bbox \nabla} -\frac{ie^* {\bf A}}{\hbar c} 
\right)^2\psi \nonumber \\
&&+ |a|\psi  -b |\psi|^2 \psi, 
\label{TDGL1}
\end{eqnarray}
\begin{equation}
{\bbox \nabla}\times {\bbox \nabla} \times {\bf A} = 
\frac{4\pi}{c}( {\bf J}_n + {\bf J}_s), 
\label{TDGL2}
\end{equation}
where the normal current ${\bf J}_n$ and the supercurrent ${\bf J}_s$
are given by 
\begin{mathletters}
\begin{eqnarray}
\label{TDGL3}
&&{\bf J}_n =   \sigma^{(n)}\left( - \partial_t {\bf A}/c
 - {\bbox \nabla} \Phi\right), \\
\label{TDGL4}
&&{\bf J}_s = {\hbar e^* \over 2 m i} \left( \psi^* {\bbox \nabla} 
\psi - \psi {\bbox \nabla} \psi^* \right) 
-\frac{e^{*2}}{mc}|\psi|^2 {\bf A}, 
\end{eqnarray}
\end{mathletters}
and where $\gamma$ (which is assumed to be real) is a dimensionless
quantity characterizing the relaxation time of the order parameter, 
$\sigma^{(n)}$ is the normal conductivity, and $a = a_0 (1- T/T_{c0})$. 
 From these parameters we can form two important length scales, 
the coherence length $\xi = \hbar/(2 m |a|)^{1/2}$ and the penetration
depth $\lambda = [mbc^2/4\pi (e^*)^2 |a|]^{1/2}$. 

These equations assume relaxational dynamics for the order parameter 
as well as a two-fluid description for the current.
 With somewhat restrictive assumptions, they can be derived from the
microscopic BCS theory \cite{schmid66,gorkov68}.   
 Further simplification is possible in the limit of a thin, narrow
film, that is, when the thickness is less than the coherence length,
$d<\xi$, and the width is less than the effective penetration depth
\cite{pearl}, $w\ll \lambda^2/d$. 
 In this case the current carried by the film or wire is small, and 
we needn't worry about the fields it produces.
 Therefore, Eq.~(\ref{TDGL2}) may be dropped, and we need only specify
the total current ${\bf J}={\bf J}_n + {\bf J}_s$ (subject to
${\bbox \nabla}\cdot{\bf J}=0$), along with the order-parameter
dynamics, Eq.~(\ref{TDGL1}).  
 This approximation is commonly used for superconducting wires 
\cite{ivlev84} and can be justified mathematically for superconducting 
films \cite{chapman97}. 
 In addition, we will be considering processes in the absence of an 
applied magnetic field, so that we may set ${\bf A}={\bf 0}$. 
 With these simplifications, we can now rewrite the equations in terms 
of dimensionless (primed) quantities,
\begin{eqnarray}
&\psi = \sqrt{{|a| \over b}}  \psi^{\prime}, 
\ \ \ \ \ \ &\Phi = {\hbar e^* |a| \over m b \sigma^{(n)}} 
 \mu^{\prime}, 
\nonumber \\
&x = \xi  x^{\prime}, 
\ \ \ \ \ \ &t = { m b \sigma^{(n)} \over e^{*2} |a|} 
 t^{\prime},  \nonumber \\
&J = \sqrt{{2 \over m}} {e^* |a|^{3/2} \over b } 
 J^{\prime},  \ \ \ \ & 
\end{eqnarray}
which leads to 
\begin{mathletters}
\label{scaled}
\begin{eqnarray}
\label{TDGL-scaled}
&&        u(\partial_{t^{\prime}} + i\mu^{\prime})\psi^{\prime} 
= (\nabla^{\prime 2} +1-|\psi^{\prime}|^2)\psi^{\prime},
 \\
\label{current-scaled}
&& {\bf J^{\prime}} = {\rm Im} (\psi^{\prime *}
{\bbox \nabla}^{\prime} {\psi^{\prime}}) -
{\bbox \nabla}^{\prime}  {\mu}^{\prime},
\quad {\bbox \nabla}^{\prime}\cdot {\bf J^{\prime}}=0 . 
\end{eqnarray}
\end{mathletters} 
Note that length is measured in units of coherence length
\cite{length}.
 We will drop the primes hereafter. 
 The only parameters remaining in the problem are the scaled current 
$J$ and a dimensionless material parameter $u=\tau_{\psi}/\tau_J$, 
where $\tau_{\psi}= \hbar \gamma/|a|$ is the order-parameter
relaxation time and $\tau_J=\sigma^{(n)} m b / e^{*2} |a|$ is the
current relaxation time. 
 We will treat $u$ as a phenomenological parameter and study the 
nucleation and growth process as a function of $u$.  
 The microscopic derivations of the TDGL equations  predict that 
$u=5.79$ (nonmagnetic impurities) \cite{schmid66}, and $u=12$ 
(paramagnetic impurities) \cite{gorkov68}, but small $u$ is also 
useful for modeling gapped superconductors \cite{ivlev80b}.

\section{Nucleation of the superconducting phase from the normal 
phase} 
\label{nuclei}

In the presence of an applied current the normal phase in a 
wire is linearly stable with respect to superconducting perturbations 
for {\it any} value of the current \cite{gorkov70,kulik70}.  
 The reason for this stability is that any quiescent superconducting 
fluctuation will be accelerated by the electric field, its velocity 
eventually exceeding the critical depairing velocity, resulting in 
the decay of the fluctuation.  
 The growth of the superconducting phase therefore requires a nucleus 
of sufficient size that will locally screen the electric field and 
allow the superconducting phase to continue growing; smaller nuclei 
will simply decay back to the normal phase.  
 The amplitude of the ``critical'' nucleus should decrease as the 
current approaches zero, reaching zero only at $J=0$. 
 We expect the {\it critical} nuclei to be unstable, stationary 
(but nonequilibrium) solutions of the TDGL equations, which
asymptotically approach the normal solution as $x \rightarrow
\pm \infty$.  
 These ``bump'' solutions of the TDGL equations are the subject of 
this section.
 We include here an extensive numerical study of the amplitudes and 
widths of the critical nuclei, as well as some analytical estimates 
for these quantities. 

\subsection{Numerical results}
\label{nuclei-num}

Let us start by discussing the numerical work on the critical 
nuclei.  
 For the analytic work, we often find it convenient to use the
amplitude and phase variables, i.e. $\psi=f {\rm e}^{i \theta}$; but
they are ill-suited for the numerical work, since the calculation of 
the phase becomes difficult when the amplitude is small. 
 Following Likharev \cite{likharev74} we use instead $\psi=R+iI$, with 
$R$ and $I$ real, and in one dimension Eqs.~(\ref{scaled}) become
\begin{mathletters}
\label{numerics}
\begin{eqnarray}
u R_t &=& R_{xx} + u\mu I + R - (R^2 + I^2)R,  \\
u I_t &=& I_{xx} - u\mu R + I - (R^2 + I^2)I,  \\
J &=& R I_x - I R_x - \mu_x.
\end{eqnarray}
\end{mathletters}

\begin{figure}
\centerline{
\epsfxsize=8.5cm
\leavevmode \epsfbox{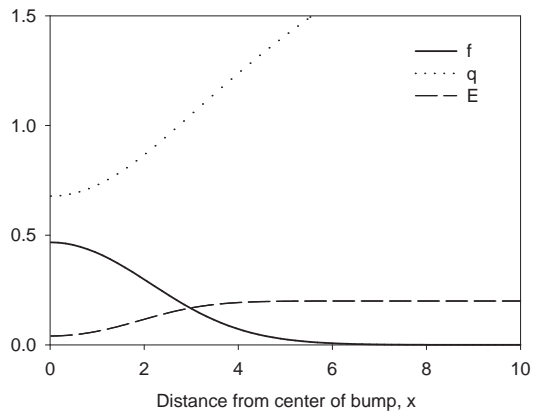}}
\caption{The bump's amplitude, $f(x)$, its 
superfluid velocity, $q(x)$, and the electric field, $E(x)$,
for $u=5.79$ and $J=0.2$. }
\label{nuc-1}
\end{figure}

\begin{figure}
\centerline{
\epsfxsize=8.5cm
\leavevmode \epsfbox{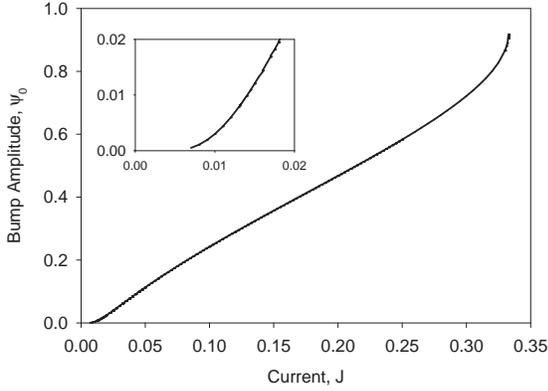}}
\caption{ The maximum amplitude of the bumps $\psi_0$ as a function
of $J$ for $u=5.79$. The inset shows the exponential dependence of
the small-$J$ data, see Eq.~(\ref{expon-depend}). }
\label{nuc-2}
\end{figure}

Since the nuclei are unstable stationary states, they are 
investigated only by time-independent means. 
 Such solutions require a particular gauge choice---in this case
$\mu(x)=0$ where $\psi(x)$ has its maximum amplitude; they are 
then sought using a relaxation algorithm \cite{num-rec}.  
 Figure~\ref{nuc-1} shows a typical bump's amplitude, $f=
\sqrt{R^2+I^2}$, the associated superfluid velocity $q=
(RI_x-IR_x)/f^2$ and the electric field $E(x)=-\mu_x(x)$. 
 The figure shows only half of the solution; $f(x)$, $q(x)$ 
and $E(x)$ are even about $x=0$. 
 In Fig.~\ref{nuc-2} we plot the bump's maximum amplitude, 
$\psi_0$, as a function of $J$; it grows as the current 
rises, indicating the increasing stability of the normal phase.
 In the data presented by Watts-Tobin {\it et al.}\,\cite{watts-tobin81} 
$\psi_0$ appears to vary linearly with $J$ for small $J$. 
 However, in our numerical calculations at very small currents the
dependence deviates from linearity (see the inset of Fig.~\ref{nuc-2}),
and $\psi_0$ drops rapidly to zero as $J\rightarrow 0$, consistent
with the exponential behavior suggested in
Refs.~\cite{ivlev84,ivlev80}.  
More precisely our small-$J$ data ($0.008\leq J\leq 0.015$) at
$u=5.79$ is fit by 
\begin{equation}
\label{expon-depend}
\psi_0(J) = B \exp ( - A/uJ),  
\end{equation}
with $A=0.042$ and $B=0.19$.
A somewhat similar dependence (with $A=2/3$) was suggested by Ivlev
{\it et al.} \cite{ivlev84,ivlev80}; they were considering a distinct 
quantity but one also related to critical fluctuations about the
normal phase (see the Appendix for more details).  


\begin{figure}
\centerline{
\epsfxsize=8.5cm
\leavevmode \epsfbox{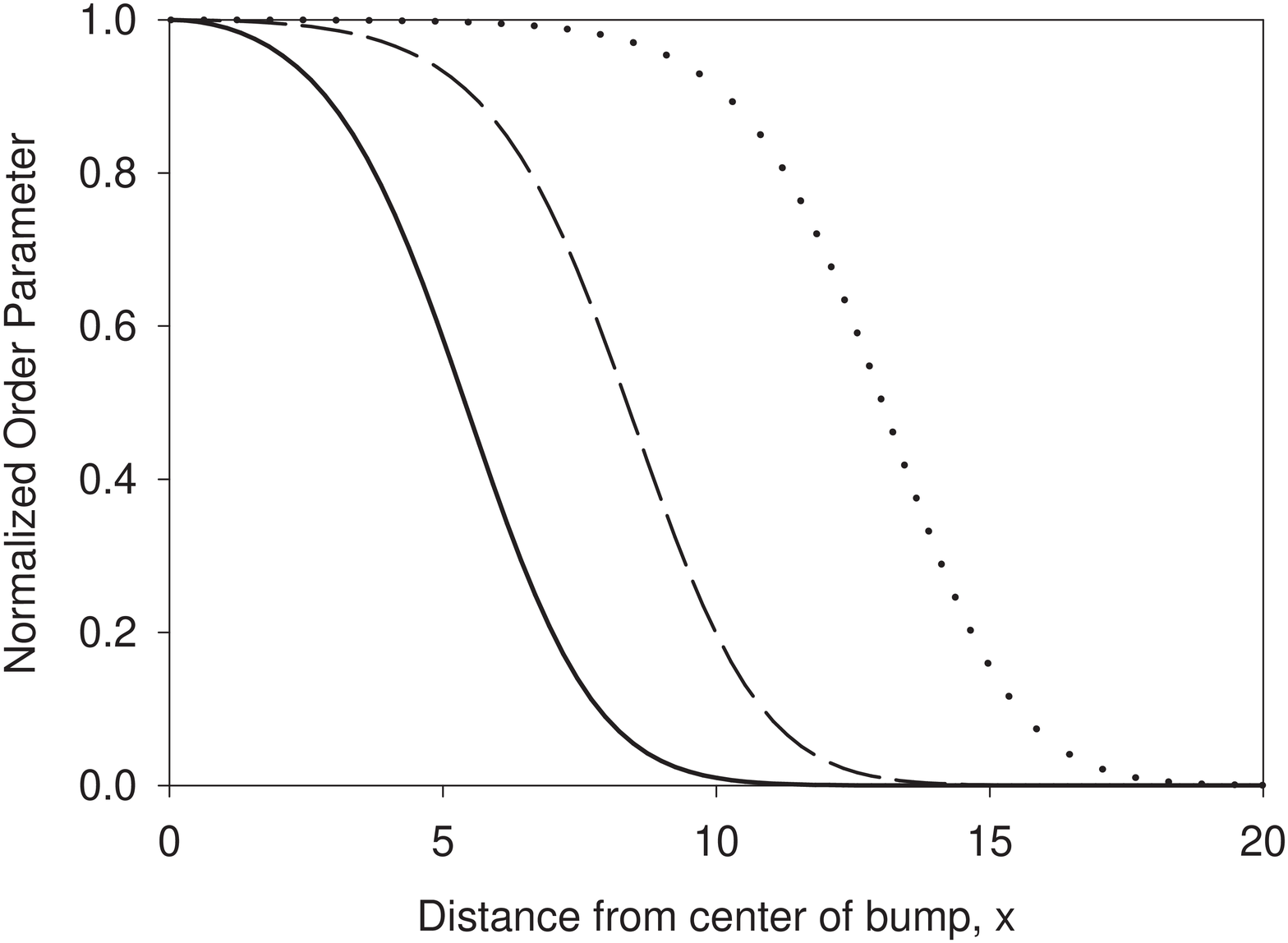}}
\caption{ The bump profiles for $J^*-J=10^{-3}$ (solid), 
$J^*-J=10^{-5}$ (dashed) and $J^*-J=10^{-7}$ (dotted) at $u=5.79$. }
\label{nuc-3}
\end{figure}

The width of the bump diverges in the small-$J$ limit like 
$(uJ)^{-1/2}$, as can be seen from the analysis below.    
 So as $J$ is increased from zero, the width initially shrinks, but 
eventually the width begins to grow again, diverging as the current 
approaches the stall current $J^*$. 
 In this limit the bump transforms into two well separated
interfaces (see Fig.~\ref{nuc-3}).

\subsection{Analysis in the $J\rightarrow 0$ limit}
\label{nuc-smallj}

The equations for nuclei centered at the origin are 
\begin{mathletters}  
\begin{eqnarray}
\label{lump-eq}
&& \psi_{xx} -iu \mu \psi +\psi -|\psi|^2\psi =0,  \\
&& \mu = -Jx + \int_0^x {\rm Im} \left(\psi^* \psi_{x'} \right) 
dx' , 
\end{eqnarray} 
\end{mathletters}
where we have dropped the term $\psi_t$ and selected the gauge
$\mu(0)=0$. 
 We saw in Fig.~\ref{nuc-2} that $\psi_0$ becomes very tiny  
in the small-$J$ limit, thus the nonlinear terms can be neglected,
leading to 
\begin{equation}
\psi_{xx} +  iuJx\psi + \psi = 0, 
\label{1}
\end{equation} 
a complex version of the Airy equation. 
 Applying the WKB method results in the approximate solution 
\begin{eqnarray}
\psi \sim &&[1+(uJx)^2]^{-1/8} \nonumber \\
\times &&\exp\left\{ \frac{2}{ 3\,uJ} \left[
\left[1 + (uJx)^2 \right]^{3/4} \cos \frac{3\alpha}{2}  - 
1 \right] \right\} \nonumber \\ 
\times &&\exp \left\{i\left[ \frac{2}{3\, uJ} 
\left[1 + (uJx)^2\right]^{3/4}\sin \frac{3\alpha}{2}
-\frac{\alpha}{4}\right] \right\},
\end{eqnarray} 
where $\alpha = \tan^{-1}(uJx)$. 
 The numerical data agrees quite well with this predicted shape in
the small-$J$ limit. 
 For small $x$ the expression can be approximated by 
\begin{equation}
\psi\sim \exp \left[i(1-uJ/4)x - uJx^2/4 \right]. 
\label{4}
\end{equation} 
We see here that the width of the bump varies like $(uJ)^{-1/2}$ 
in this limit and that the superfluid velocity 
$q \approx (1-uJ/4)$.  
 For large $x$, on the other hand, where $\alpha \approx \pi/2$, 
the expression becomes 
\begin{equation}
\label{airy}
\psi\sim (uJx)^{-1/4} \exp \left[ - 
{\sqrt{2uJ}\over 3} |x|^{3/2}(1-i)\right],
\label{5}
\end{equation} 
as one expects for the Airy function.  
 Note that deep in the tail of the solution, we see a different
length scale $\lambda_{Airy} \sim (uJ)^{-1/3}$ arising.  

Since the above analysis is of a linear equation, it can 
not determine the amplitude of the nucleus;  for this purpose the
nonlinearities must be considered.  
 In the appendix we outline an {\it ad hoc} calculation of the
small-$J$ limit of the bump amplitude.
 We take a $\psi$ of an unknown amplitude but of a fixed shape
inspired by the above analysis and assume that it is a stationary
solution of the full TDGL.
 We then determine its amplitude self-consistently.  
 The resulting amplitude is 
\begin{equation}
\label{self}
|\psi| \approx  
\left(\frac{2J}{\pi u}\right)^{1/4} 
\left(\frac{9}{8}  - \frac{1}{u} \right)^{-1/2}
\exp \left(- \frac{16}{81~uJ} \right). 
\end{equation}
The factor $A=16/81$ is within a few percent of that 
extracted from the numerical data.

\section{Stationary Interfaces}
\label{stationary}

As the current is raised, the width of the critical nucleus 
grows and ultimately diverges as the stall current is reached,
resulting in well separated, stationary interfaces. 
These interface solutions will be the subject of the rest of this
work. 

\subsection{Numerical methods and results} 
\label{stat-num}

Let us first discuss the numerical work on the interface 
solutions. 
 For given values of $u$ and $J$ we evolved the TDGL equations from an
initial guess which is purely superconducting on the left, $\psi(x)=
f_{\infty}~{\rm e}^{iq_{\infty}x}$ and  $\mu_x(x)=0$, and purely 
normal on the right, $\psi(x)=0$ and $\mu_x(x)=-J$.
 The values $f_{\infty}$ and $q_{\infty}$ are related to the 
applied current through 
\begin{mathletters}
\label{boundary-relations}
\begin{eqnarray}
J &=& f_{\infty}^2 \sqrt{1-f_{\infty}^2}, \\
q_{\infty} &=& \sqrt{1-f_{\infty}^2}.  
\end{eqnarray} 
\end{mathletters} 
Stability requires taking the larger positive root of the former 
equation \cite{langer67} which places the following bounds on $J$, 
$f_{\infty}$ and $q_{\infty}$:
\begin{mathletters}
\label{bounds}
\begin{eqnarray} 
&&0 \leq J \leq J_c = \sqrt{4/27} \approx 0.3849, 
\label{j-bound} \\
&&1 \geq f_{\infty} \geq \sqrt{2/3} \approx 0.8165,
\label{f-bound} \\ 
&&0 \leq q_{\infty} \leq \sqrt{1/3} \approx 0.5774.
\label{q-bound}
\end{eqnarray}
\end{mathletters}   
We employed several schemes to integrate the equations in time
including both explicit (Euler) and implicit (Crank-Nicholson) 
\cite{num-rec}.
    
Initially the front moves and changes shape but eventually it reaches
a steady state in which the interface moves at a constant velocity
without further deformation.
 By the time-dependent means we found locally stable,
constant-velocity solutions for currents less than $J_c$.
 To examine these solutions more accurately, we adopted a 
time-independent method. 
 First, we transformed coordinates to a moving frame, $x^{\prime}=
x-ct$; next, we chose $\mu=cq_{\infty}$ as $x\rightarrow -\infty$
which allows for a truly time-independent solution (i.e.\ one with
both amplitude and phase time-independent).
 Then we searched for stationary solutions using a relaxation
algorithm \cite{num-rec} where ($u,J$) are input parameters and $c$ is
treated as an eigenvalue. 
 This approach requires an additional boundary condition to fix
translational invariance; we elected to fix $\mu$ on the rightmost
site.
 To find the stall currents $J^*$ we can set $c=0$ and take $J$ or $u$
as the eigenvalue. 

\begin{figure}
\centerline{
\epsfxsize=8.5cm
\leavevmode \epsfbox{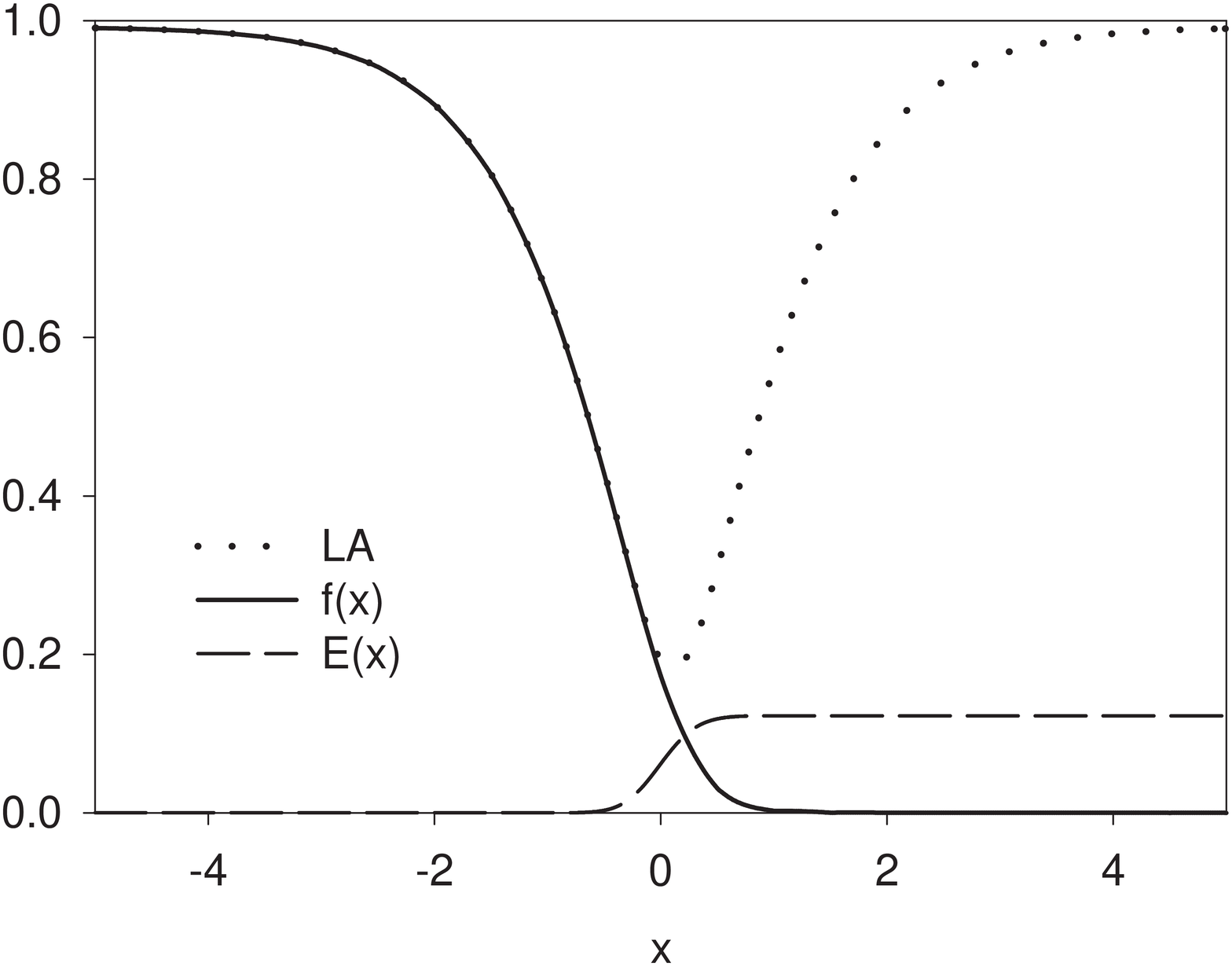}}
\caption{The stationary NS interface solution when $u=500$
for which the stall current $J^*=0.12252$. 
 Shown here are the numerically determined $f(x)$ and 
$E(x)$, as well as the Langer-Ambegaokar (LA)
solution (Eq.~(\ref{langer}), the solution with no electric
field) corresponding to the same current.}
\label{fig1}
\end{figure}

\begin{figure}
\centerline{
\epsfxsize=8.5cm
\leavevmode \epsfbox{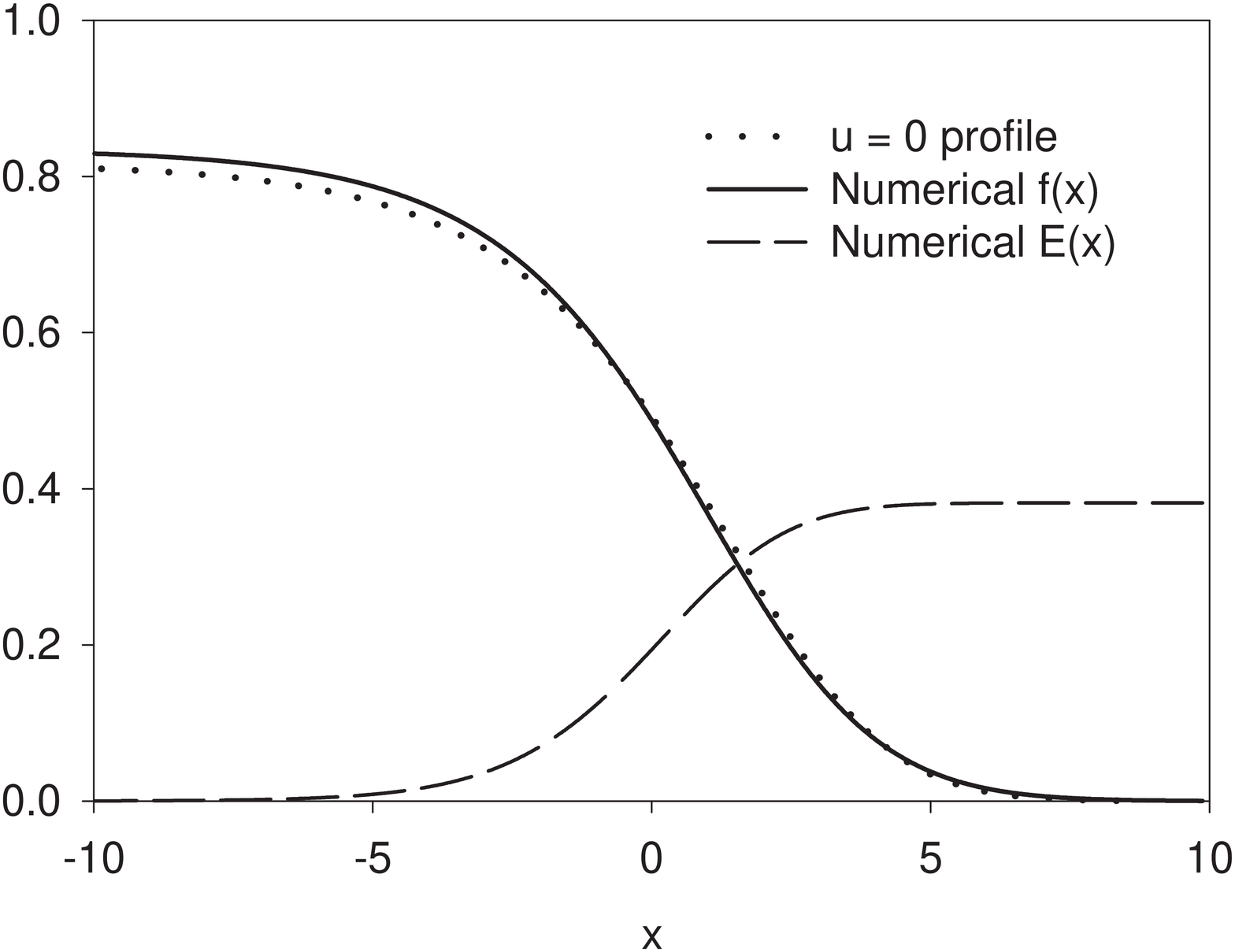}}
\caption{ The stationary NS interface solution when $u=1.04$ for
which the stall current $J^*=0.3836$. 
 Shown here are the numerically determined $f(x)$ and $E(x)$, as
well as the function ${\hat f}_0(\hat x)$, the $u \rightarrow 0$
profile, derived from Eq.~(\ref{profile-inverse}) where
$\hat x= u^{1/2}x$.}
\label{fig2}
\end{figure}

Figures \ref{fig1} and \ref{fig2} show the order-parameter amplitude
$f$ and the electric-field distribution $E=-\mu_x$ of the stalled
interface determined for $u=500$ and $u=1.04$ respectively. 
 Note that while $f$ is very flat in the superconducting region, the
real and imaginary parts, $R$ and $I$, oscillate with a wavelength
$2 \pi /q_{\infty}$.
 Because of this additional length scale inherent in $R$ and $I$,  
there is little to be gained from varying the mesh size.
 In fact, this length scale is compressed as we move to the right,
and we are only saved from the difficulties of handling rapidly
oscillating functions by the fact that the amplitudes decay so
quickly.

In the large-$u$ case (see Fig.~\ref{fig1}), $E(x)$ remains flat 
throughout most of the space; it changes abruptly from one constant 
to another only after $f(x)$ has become small. 
 The variations of $f(x)$ are more gradual; however, the 
greatest changes in $f_x$ occur in that same small area.  
 This region of rapid change is known as a {\it boundary layer}; 
it marks where the current suddenly changes from superconducting to
normal, i.e., the position of the NS interface. 
 As $u$ increases, the longer length scale over which $f$ varies 
on the superconducting side remains essentially fixed, while 
the boundary-layer thickness shrinks to zero. 
 In the opposite limit, the small-$u$ case (see Fig.~\ref{fig2}), 
$f(x)$ and $E(x)$ appear to vary together even in the superconducting
region; moreover, the length scale over which they vary grows as $u$
is decreased. 
 We will postpone providing more of the numerical results on the
interfaces until some of the analytic arguments are available
for comparison. 

\subsection{Asymptotic analysis of the interface solutions: 
preliminaries}
\label{stat-prelim}

Before addressing the large-$u$ and small-$u$ limits separately,
let us put the TDGL equations into a form convenient for analysis
and derive expressions for the length scales deep in the
superconducting and normal regions. 
 The disparity of these length scales in the large-$u$ limit will 
motivate the boundary-layer analysis in that regime; while an 
inequality they satisfy will lead to the conclusion that 
$J^* \rightarrow J_c$ in the small-$u$ limit. 

We make the substitution $\psi = f e^{i\theta}$, which
yields 
\begin{mathletters}
\label{analytic}
\begin{eqnarray}
&& uf_t = f_{xx} - f (\theta_x)^2   +f - f^3 ,
\label{analytic-1} \\
&&u(\theta_t+\mu )f= 2f_x\theta_x + f \theta_{xx},
\label{analytic-2} \\
&& J =  f^2 \theta_x - \mu_x.
\label{analtyic-3}
\end{eqnarray}
\end{mathletters} 
Next we restrict our attention to stationary solutions.
 Note that only spatial derivatives of $\theta$ appear now, allowing
us to work with the superfluid velocity $q=\theta_x$ instead of
$\theta$.
 The equations become
\begin{mathletters}
\label{fandq}
\begin{eqnarray}
&&  f_{xx} - q^2 f  +f - f^3=0 ,
\label{fandq-1} \\
&& u \mu f= 2f_x q + f q_{x},
\label{fandq-2} \\
&& J^* =  f^2 q - \mu_x,
\label{fandq-3}
\end{eqnarray}
\end{mathletters}
where $J^*$ replaces $J$ as these equations apply to the stall
situation. 
 Next multiply Eq.~(\ref{fandq-2}) by $f$ and note that the right
hand side is now $(f^2q)_x$ which we can express in terms of $\mu$
by differentiating Eq.~(\ref{fandq-3}); these steps lead to
\begin{equation}
\mu_{xx}=u f^2 \mu .
\label{fandmu}
\end{equation}

Now let us assume the following asymptotic forms as 
$x\rightarrow -\infty$:  
\begin{mathletters}
\label{series}
\begin{eqnarray}
\label{series-f}
\lim_{x \to -\infty} f(x) &=& f_{\infty} - 
f_1 ~{\rm e}^{x/\lambda_{f}} + \ldots , \\
\label{series-q}
\lim_{x \to -\infty} q(x) &=& q_{\infty} + 
q_1 ~{\rm e}^{x/\lambda_{q}} + \ldots , \\
\label{series-mu}
\lim_{x \to -\infty} \mu(x) &=& \mu_{\infty} - 
\mu_1 ~{\rm e}^{x/\lambda_{\mu}} + \ldots.   
\end{eqnarray}
\end{mathletters}  
Substituting these expressions into Eqs.~(\ref{fandq-1}), 
(\ref{fandq-3}) and (\ref{fandmu}) and recalling the boundary
conditions yields 
\begin{mathletters}
\label{approach}
\begin{eqnarray}
\label{approach-1}
&&\left(2f_{\infty}^2-\lambda_f^{-2} \right)
f_1 {\rm e}^{x/\lambda_f} -2 f_{\infty} q_{\infty}q_1 
{\rm e}^{x/\lambda_q} =0,
\\ 
\label{approach-2} 
&&-2f_{\infty} q_{\infty} f_1 {\rm e}^{x/\lambda_f}+
f_{\infty}^2q_1 {\rm e}^{x/\lambda_q} +
\lambda_{\mu}^{-1} \mu_1 {\rm e}^{x/\lambda_{\mu}}=0, \\ 
\label{approach-3} 
&&\lambda_{\mu}^{-2} - u f_{\infty}^2=0.
\end{eqnarray}
\end{mathletters} 
Eq.~(\ref{approach-3}) provides an expression for $\lambda_{\mu}$,
the electric-field screening length.
 Since $f_{\infty}$ is always of $O(1)$, we see that
$\lambda_{\mu}$ shrinks as $u \rightarrow \infty$ and diverges as
$u \rightarrow 0$, which is consistent with the behavior seen in
Figs.~\ref{fig1} and \ref{fig2}. 

 
More than one decay length appears in  Eqs.~(\ref{approach-1}) and
(\ref{approach-2}). 
 If they are not equal, the term with the shorter length is
exponentially small compared to the other(s) and will not
contribute to the $x \rightarrow -\infty$ limit.     
 Since none of the terms in Eq.~(\ref{approach-2}) can equal
zero individually, we conclude that the longer two of $\lambda_f$,
$\lambda_q$ and $\lambda_{\mu}$ must be equal.   
 Next, because the term multiplying ${\rm e}^{x/\lambda_q}$ in
Eq.~(\ref{approach-1}) cannot equal zero on its own, we determine
that $\lambda_q \leq \lambda_f$, making $\lambda_f$ one of the longer
lengths.  
 Finally, if we assume that $\lambda_f=\lambda_{\mu} > \lambda_q$ we 
find that $\lambda_f=2^{-1/2}f_{\infty}^{-1}$ and $\lambda_{\mu}=
u^{-1/2}f_{\infty}^{-1}$ and reach a contradiction (except at $u=2$).  
 Thus provided the original assumption of an exponential approach is 
valid, we conclude that 
\begin{equation}
\label{inequality}
\lambda_f=\lambda_q \geq \lambda_{\mu}. 
\end{equation}
This equality of $\lambda_f$ and $\lambda_q$ is reasonable given that
both $f$ and $q$ are related to the complex order parameter $\psi$. 
 Also, having $\lambda_f > \lambda_{\mu}$ is consistent with the 
large-$u$ data seen in Fig.~\ref{fig1}.   
 If $\lambda_{\mu} \ne \lambda_f$ then 
\begin{equation}
\label{LA-length}
\lambda_f^{-2} = 6 f_{\infty}^2 -4 = \lambda_{LA}^{-2}. 
\end{equation}
We identify this length scale as $\lambda_{LA}$ since it
coincides with that occurring in the solution of Eqs.~(\ref{fandq})
without any electric field ($\mu(x)=0$), 
\begin{equation}
\label{langer}
f^2(x)=   f_{\infty}^2 - (3f_{\infty}^2 -2)
~{\rm sech}^2 \left(  \sqrt{\frac{3f_{\infty}^2-2}{2}}
x \right)  ,
\end{equation}
which was found by Langer and Ambegaokar \cite{langer67} in their 
study of phase slippage.
 The asymptotic form of Eq.~(\ref{langer}) looks like
Eq.~(\ref{series-f}) with $\lambda_f$ given by Eq.~(\ref{LA-length}). 
 As a matter of fact because $\lambda_f \gg \lambda_{\mu}$ in the 
large-$u$ limit, the profile of $f(x)$ is only imperceptibly different 
from the Langer-Ambegaokar (LA) solution in the superconducting
region and deviates from it only in the boundary layer, as is shown
in Fig.~\ref{fig1}.  

Recall that $\lambda_{\mu}$ diverges as $u \rightarrow 0$; the
inequality $\lambda_f \geq \lambda_{\mu}$ implies that $\lambda_f$
must diverge as fast or faster in this limit. 
 This scenario is consistent with the small-$u$ data shown in 
Fig.~\ref{fig2} in which $f(x)$ and $E(x)$ vary on long length
scales.  
 Eq.~(\ref{LA-length}) suggests that a diverging $\lambda_f$ implies
that $f_{\infty} \rightarrow \sqrt{2/3}$ and in turn that
$J \rightarrow J_c$ as $u \rightarrow 0$, which is also consistent
with what is found numerically.  

In the other asymptotic limit, deep in the normal regime, $\psi$ is
very small and hence the nonlinear terms in Eqs.~(\ref{scaled}) can
be dropped as was done for the bumps in the small-$J$ limit. 
 The result is a complex Airy equation, the asymptotic analysis of
which was supplied in Eq.~(\ref{airy}), where we saw the length
scale $\lambda_{Airy} \sim (u J^*)^{-1/3}$.  
 Somewhat like $\lambda_{\mu}$, $\lambda_{Airy}$ shrinks as
$u \rightarrow \infty$ and expands as $u \rightarrow 0$ but with
different powers of $u$.
 The presence of the disparate length scales, $\lambda_f$, 
$\lambda_{\mu}$ and $\lambda_{Airy}$, in the large-$u$ limit 
motivates the use of the boundary-layer analysis that comes next. 
 We will see that $\lambda_{Airy}$ scales in the same way as
the boundary-layer thickness.

\subsection{Asymptotic behavior of the stall current as 
$u\rightarrow\infty$}
\label{stat-largeu}

We have already seen in Fig.~\ref{fig1} that the large-$u$
profile can be divided into two regions---one slowly varying,
one rapidly varying, also known as the {\it outer} and {\it inner}
regions, respectively. 
 Furthermore, it has been suggested that the ratio of the length
scales characterizing these regions decreases as $u \rightarrow
\infty$. 
 These features make the problem ideally suited for
boundary-layer analysis, in which one identifies the terms that
dominate the differential equation in each region, analyzes the
reduced equations consisting of dominant terms and then matches
the behavior in some intermediate region. 
 
We start by eliminating the superfluid velocity $q$  from
Eqs.~(\ref{fandq}), resulting in 
\begin{mathletters}
\begin{eqnarray}
\label{feqn-1}
&& f_{xx} - (J^* + \mu_x)^2  f^{-3}   +f - f^3 = 0, \\
\label{mueqn-1}
&& \mu_{xx} - u f^2\mu = 0. 
\end{eqnarray}
\end{mathletters}
Let us consider first the slowly varying, superconducting region.
 We saw in the preliminary analysis that for large $u$, $\mu(x)$
is exponentially small, so we drop it. 
 Next, let us assume that $J^*$ is small and drop it; we can verify 
in the end that this is self-consistent.  
 The reduced equation is 
\begin{equation}
\label{outer}
f_{xx}+f-f^3\approx 0,
\end{equation}
with solution $f(x)=-{\rm tanh}(x/\sqrt{2})$. 

Moving in from the left toward the interface (into the
boundary-layer region), $f$ becomes small, and the second
term in Eq.~(\ref{feqn-1}) which was subdominant becomes dominant.
 In this inner region $f$ is small but rapidly varying, thus the
dominant terms are
\begin{equation}
\label{reduced-f}
f_{xx} \approx \frac{(J^*+\mu_x)^2}{f^3},
\end{equation}
along with Eq.~(\ref{mueqn-1}).
 Having identified the dominant terms, now we must make certain
they balance.
 We assume that in the boundary layer, all the quantities scale
as powers of $u$: 
\begin{equation}
f \sim u^{-\alpha},  \quad \mu \sim u^{-\beta}, \quad
J^* \sim u^{-\gamma},  \quad x \sim u^{-\delta}.   
\end{equation}
Balancing terms in Eq.~(\ref{reduced-f}), we find 
$2\alpha = \gamma + \delta$, while balancing terms in
Eq.~(\ref{mueqn-1}) yields $2(\alpha + \delta) = 1$.
 Next, we need to ensure that the solutions in the boundary layer
match onto the solutions in the superconducting and normal regions.
 By expanding the superconducting solution near the interface, we see
that $f(x)\sim - x/\sqrt{2}$ as the boundary layer is approached;
matching to the boundary layer requires $f_x\sim 1$, so that
$\alpha = \gamma$.
 In the normal region, $\mu \approx - J^* x$, so that matching to the
boundary layer requires $\mu_x \sim J^*$, and $\beta = \gamma +
\delta$.
 Solving this set of equations, we conclude that $\alpha=\gamma=
\delta=1/4$ and $\beta=1/2$, i.e., the stall current $J^* \sim
u^{-1/4}$ for large $u$. 
 Note that $J^* \rightarrow 0$ as $u \rightarrow \infty$, so
that we were justified in dropping $J^{*2}/f^3$ from
Eq.~(\ref{outer}). 
 Substituting $J^* \sim u^{-1/4}$ into $\lambda_{Airy}$ gives 
$\lambda_{Airy} \sim u^{-1/4}$, indicating that it may be
identified as the boundary-layer thickness.   
    
\begin{figure}
\centerline{
\epsfxsize=8.5cm
\leavevmode \epsfbox{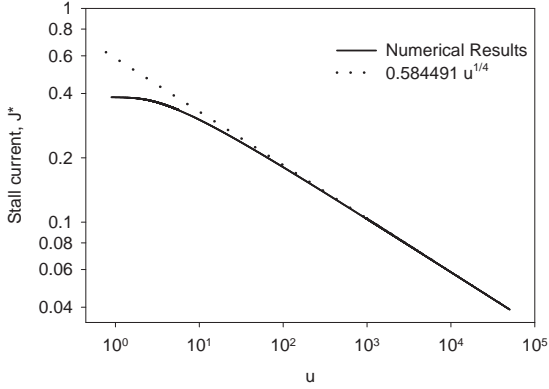}}
\caption{A log-log plot of the stall current $J^*$ vs. $u$.  The 
solid line shows the numerically determined $J^*$'s as a function 
of $u$ and the dotted line is $0.584491 ~u^{-1/4}$ (the large-$u$
behavior predicted by matched asymptotic analysis).} 
\label{drew3}
\end{figure}

In order to determine the coefficient of the $u^{-1/4}$ in the 
stall current we need to solve the boundary layer problem. 
 Let us rescale in the way suggested above: 
\begin{eqnarray}
&&f = u^{-1/4}F,\quad \mu = u^{-1/2} M(X), \nonumber \\
&&J = u^{-1/4} \tilde J, \quad x=u^{-1/4} X. 
\label{inner1}
\end{eqnarray}
Substituting these rescaled variables into Eqs.~(\ref{feqn-1}) and 
(\ref{mueqn-1}), and then expanding $F$, $M$ and $\tilde J$ in powers
of $u^{-1/2}$, we obtain at the lowest order  
\begin{mathletters}
\begin{eqnarray}
\label{inner2}
&&F_{0,XX} - \frac{({\tilde J}_0 + M_{0,X})^2}{ F_0^3}  =0,  \\ 
\label{inner3}
&&M_{0,XX} -  F_0^2 M_0 = 0,  
\end{eqnarray}
\end{mathletters}
with the boundary conditions (from the outer regions) 
\begin{eqnarray}
F_{0,X}(-\infty) & = -1/\sqrt{2},
\qquad M_0(-\infty) & = 0, \\
F_0(+\infty) & = 0,
\qquad M_{0,X}(+\infty) & = -{\tilde J}_0.
\label{inner_boundary}
\end{eqnarray} 
(As before we need an extra boundary condition to fix the
translational invariance.) 
 For an arbitrary ${\tilde J}_0$ the solutions of Eqs.~(\ref{inner2})
and (\ref{inner3}) cannot satisfy the boundary conditions;
${\tilde J}_0$ must be tuned to a particular value before all of the
boundary conditions are satisfied, leading to a {\it nonlinear 
eigenvalue problem} for ${\tilde J}_0$. 
 We have solved this eigenvalue problem numerically and find that
${\tilde J}_0= 0.584491$. 
 Therefore, to leading order we have for the stall current  
\begin{equation}
J^* \approx 0.584491\, u^{-1/4}.
\label{asympt}
\end{equation}
This prediction agrees well with the numerical results and
disagrees with Likharev's conjecture of a $u^{-1/3}$ dependence
\cite{likharev74}, as can be seen in Fig.~\ref{drew3} and 
in Table~\ref{table2}. 
 It is in principle possible to carry out this procedure to 
successively higher orders, but the equations become cumbersome. 
 Instead we have simply opted to fit our numerical data to a form
inspired by the asymptotic analysis, 
\begin{eqnarray}
J^* &=& 0.584491\, u^{-1/4} -  0.117461\, u^{-3/4}  
- 0.12498\, u^{-5/4} \nonumber \\
&&+ 0.163043\, u^{-7/4} 
+O\left(u^{-9/4}\right). 
\end{eqnarray} 

\begin{table}
\caption{Representative numerical results for the stall
current $J^*$ and kinetic coefficient $\eta$.}
\label{table2}
\begin{tabular}{lllll}
 $u $ & $J^*$   & $ J^* u^{1/4} $ & $ \eta $ & $\eta u^{-3/4}$ \\
\tableline
1     & 0.3838  & 0.3838   & 0.01871  & 0.01871 \\
5     & 0.3407  & 0.5094   & 0.6400  & 0.1914 \\
10    & 0.3013  & 0.5359   & 1.573  & 0.2797 \\
50    & 0.2127  & 0.5655   & 8.258 & 0.4315 \\
100   & 0.1807  & 0.5715   & 15.59  & 0.4931 \\
500   & 0.1224  & 0.5788   & 62.51  & 0.5875 \\
1000  & 0.1033  & 0.5807   & 111.3 & 0.6259 \\
5000  & 0.0693  & 0.5828   & 407.9  & 0.6847 \\
10000 & 0.0583  & 0.5833   & 708.4  & 0.7084 \\
50000 & 0.0391  & 0.5840   & 2487  &  0.7440 
\end{tabular}
\end{table}

\subsection{Asymptotic behavior of the stall current as 
$u\rightarrow 0$}
\label{stat-smallu}

Now let us examine the opposite limit of $u \rightarrow 0$. 
 In this case the electric-field screening length becomes long, and  
Ivlev {\it et al.} \cite{ivlev80b} have proposed that this makes 
the small-$u$ limit useful for modeling gapped superconductors. 
 As already suggested the inequality of length scales, $\lambda_f
\geq \lambda_{\mu}$ implies that $J^* \rightarrow J_c$. 
 We will begin our small-$u$ analysis by putting this result 
on firmer ground and extracting as a byproduct the $u \rightarrow
0$ limit of the interface profile.    

{\it The rescaled equations.} 
Recall that deep in the superconducting region $\lambda_{\mu} \sim
u^{-1/2}$. 
 This observation suggests that we rescale distance:
$x = u^{-1/2}\hat x$; furthermore, to ensure that the normal current
($-\mu_x$) scales in the same way as the total current we rescale
$\mu = u^{-1/2}\hat \mu$ as well. 
 These rescalings yield 
\begin{mathletters}
\begin{eqnarray}
\label{outer1}
&& u \hat f_{\hat x \hat x} - {\hat q}^2 \hat f +
\hat f -{\hat f}^3 = 0, \\
\label{outer2}
&& \hat \mu \hat f = 2 \hat q {\hat f}_{\hat x} +
\hat f {\hat q}_{\hat x}, \\
\label{outer3}
&& J^* = {\hat f}^2 \hat q - {\hat \mu}_{\hat x},  
\end{eqnarray}
\end{mathletters}
placing the small parameter $u$ in front of ${\hat f}_{\hat x
\hat x}$. 
 If we expand these functions as series in powers of $u$ 
\begin{mathletters}
\begin{eqnarray}
\label{expand1}
\hat f &=& {\hat f}_0 + u{\hat f}_1  + \ldots, \\
\label{expand2}
\hat q &=& {\hat q}_0 + u {\hat q}_1  + \ldots, \\
\label{expand3}
\hat \mu &=& {\hat \mu}_0 + u{\hat \mu}_1  + \ldots, \\
\label{expand4}
J^* &=& J_0^* + uJ_1^*  + \ldots,  
\end{eqnarray}
\end{mathletters}
then we find at the lowest order 
\begin{mathletters} 
\begin{eqnarray} 
\label{order11} 
&&-{\hat q}_0^2 {\hat f}_0 + {\hat f}_0 - {\hat f}_0^3 = 0, \\
\label{order12} 
&&{\hat \mu}_0 {\hat f}_0 = 2 {\hat q}_0 {\hat f}_{0,x} +
{\hat f}_0 {\hat q}_{0,x}, \\ 
\label{order13}
&&J_0^* = {\hat f}_0^2 {\hat q}_0 - {\hat \mu}_{0,\hat x}. 
\end{eqnarray}
\end{mathletters}

The solution of Eq.~(\ref{order11}) is either ${\hat f}_0 = 0$ (the
normal phase) or ${\hat f}_0 = (1-{\hat q}_0^2)^{1/2}$ (the
superconducting phase). 
 Let us focus on the superconducting solutions.  
 By eliminating ${\hat q}_0$, we obtain the first order equations 
\begin{mathletters} 
\begin{eqnarray} 
\label{order14} 
{\hat f}_{0,\hat x} &=& \frac{{\hat f}_0 \sqrt{1-{\hat f}_0^2}
~{\hat \mu}_0}{2-3 {\hat f}_0^2}, \\
\label{order15}
{\hat \mu}_{0,\hat x} &=& {\hat f}_0^2 \sqrt{ 1 - {\hat f}_0^2} -
J_0^*. 
\end{eqnarray}
\end{mathletters} 
Because ${\hat f}_0$ ranges from $f_{\infty}$ to $0$ and $f_{\infty}
\geq \sqrt{2/3}$, we know that ${\hat f}_0$ either starts at or passes
through $\sqrt{2/3}$. 
 (Strictly speaking we should be writing here $f_{\infty,0}$, the
lowest order term in the expansion of $f_{\infty}$.) 
 Thus, the effect of the pole in Eq.~(\ref{order14}) must be
considered. 
 If it is not canceled by a zero in ${\hat \mu}_0$,
${\hat f}_{0,\hat x}$ diverges at ${\hat f}_0=\sqrt{2/3}$.  

We can obtain an expression for ${\hat \mu}_0({\hat f}_0)$ by dividing
Eq.~(\ref{order15}) by Eq.~(\ref{order14}), which leads to 
\begin{equation}
{\hat \mu}_0\, d{\hat \mu}_0 = \frac{\left[{\hat f}_0^2
\sqrt{1-{\hat f}_0^2 } - J^*_0\right]\left(2-3{\hat f}_0^2\right)}
{ {\hat f}_0 \sqrt{1-{\hat f}_0^2}} 
d{\hat f}_0.
\label{combine}
\end{equation}
Integrating both sides and recalling the boundary condition
$\mu_{\infty}=0$, we find 
\begin{eqnarray}
\frac{{\hat \mu}_0^2}{2} &=& 
{\hat f}_0^2-\frac{3}{4}\,{\hat f}_0^4- f_{\infty}^2+
\frac{3}{4}\,f_{\infty}^4
\nonumber \\
&&+2\,J^*_0 \ln \left[ \frac {1+\sqrt {1-{\hat f}_0^2}}{{\hat f}_0}
\right]-3\,J^*_0\sqrt {1-{\hat f}_0^2} 
\nonumber \\
&&-2\,J^*_0\ln \left[
\frac {1+\sqrt {1-f_{\infty}^2}}{ f_{\infty}}\right]
+3\,J^*_0\sqrt {1-f_{\infty}^2}, 
\label{mess}
\end{eqnarray}
where $J^*_0 = f_\infty^2 \sqrt{1-f_\infty^2}$. 
 To keep ${\hat f}_{0,x}$ from diverging, we insist that
${\hat \mu}_0({\hat f}_0=\sqrt{2/3})=0$ which can be shown from
Eq.~(\ref{mess}) to imply $f_{\infty}=\sqrt{2/3}$, i.e.\
the small-$u$ limit of the stall current is the critical depairing
current. 
 Note that the pole in Eq.~(\ref{order14}) and the compensating
zero in ${\hat \mu}_0({\hat f}_0)$  occur at the boundary 
($x \rightarrow -\infty$). 

We can rearrange Eq.~(\ref{order14}) as follows 
\begin{equation}
\label{profile-inverse}
\int_{{\hat f}_0(0)}^{{\hat f}_0(\hat x)} \frac{(2-3f^2)~df}
{f \sqrt{1-f^2} ~{\hat \mu}_0(f)}= {\hat x}.  
\end{equation} 
Then we can substitute in Eq.~(\ref{mess}) for ${\hat \mu}_0(f)$,
numerically integrate the resulting expression and finally invert
it in order to calculate ${\hat f}_0(\hat x)$, the $u \rightarrow 0$
profile.
 Figure \ref{fig2} includes a comparison of ${\hat f}_0(\hat x)$ and
the profile of a small-$u$ numerical solution.   

To find the asymptotic behavior of ${\hat f}_0$ and ${\hat \mu}_0$ in
the superconducting region, Taylor expand ${\hat \mu}_0({\hat f}_0)$
around $f_{\infty}$
\begin{equation}
{\hat \mu}_0({\hat f}_0) = -3 \sqrt{2} \left({\hat f}_0 -\sqrt{2/3}
\right)^2 +\ldots.
\end{equation}
Notice that ${\hat \mu}_0({\hat f}_0)$ is a second order zero, so
that ${\hat f}_{0, \hat x}=0$, as it should at the boundary. 
 As a consequence, the integral supplying the inverse profile,
Eq.~(\ref{profile-inverse}), has a pole; integrating the
expression in its neighborhood yields $\sqrt{6} \ln (\sqrt{2/3}-
{\hat f}_0)$, leading to
\begin{equation}
\label{sol1}
{\hat f}_0(\hat x)\sim \sqrt{2/3} - A_0~{\rm exp}
\left( \hat x/\sqrt{6} \right), 
\end{equation}
where $A_0$ is an integration constant undetermined because of the
translational invariance. 
 Note ${\hat f}_0(\hat x)$ has the form assumed in the preliminary
analysis with $\lambda_{f,0}=\sqrt{u/6}$. 
 Putting this result into Eq.~(\ref{order14}) leads to 
\begin{equation}
\label{sol2}
{\hat \mu}_0(\hat x)\sim  - 3\sqrt{2} A_0^2~{\rm exp}
\left( 2 \hat x/ \sqrt{6}\right),
\end{equation}
where $\lambda_{\mu,0}=u^{1/2} f_{\infty}$ in agreement with the
expression found previously.

Let us examine Eqs.~(\ref{order14}) and (\ref{order15}), which
are strictly speaking superconducting solutions, in the normal
(small-${\hat f}_0$) limit. 
 Eq.~(\ref{order15}) leads to ${\hat \mu}_0(\hat x)\approx -J_c \hat x$,
and inserting this into Eq.~(\ref{order14}) reveals that ${\hat f}_0
\rightarrow 0$ in the following way 
\begin{equation}
\label{sol3}
{\hat f}_0(\hat x)\sim {\rm exp} \left( - J_c {\hat x}^2/4 \right). 
\end{equation}
This same dependence was seen earlier in the analysis of the bump 
shapes in the small-$J$ limit, Eq.~(\ref{4}). 

What is surprising here is that what are ostensibly the ``outer'' 
equations for the superconducting region also satisfy the boundary 
conditions in the normal region and interpolate in between.  
 This is consistent with the numerical observation that there does 
not seem to be a boundary layer at small $u$, that the $u f_{xx}$ 
term is apparently {\it not} a singular perturbation. 
 With this in mind, we pursue the perturbative expansion to higher
orders.

{\it The $O(u)$ equations.} 
The eigenvalue $J^*_0$ was determined by examining the behavior deep
in the superconducting region and did not require imposing the
boundary conditions on the normal side.
 Furthermore, the spatial dependence of the solution in this
region is of the form assumed in Eqs.~(\ref{series}).
 We exploit these features to obtain higher order terms.  
 The $O(u)$ equations are   
\begin{mathletters}
\label{order-u}
\begin{eqnarray}
&& {\hat f}_{0, \hat x \hat x} - 2 {\hat q}_0 {\hat f}_0
{\hat q}_1 - {\hat q}_0^2 {\hat f_1} - 3 {\hat f}_0^2 {\hat f}_1 = 0, 
\\
&& {\hat \mu}_0 {\hat f}_1 + {\hat f}_0 {\hat \mu}_1 =
2 {\hat q}_0 {\hat f}_{1, \hat x} + 2 {\hat f}_{0, \hat x}
{\hat q}_1 + {\hat f}_0 {\hat q}_{1, \hat x} + {\hat q}_{0, \hat x}
{\hat f}_1 \\
&& J^*_1 = 2 {\hat f}_0 {\hat q}_0 {\hat f}_1 + {\hat f}_0^2
{\hat q}_1 - {\hat \mu}_{1, \hat x}. 
\end{eqnarray}
\end{mathletters}
The asymptotic form of ${\hat f}_0(x)$ is
\begin{equation}
{\hat f}_0(x) = {\hat f}_0^{(0)}+{\hat f}_0^{(1)}~{\rm e}^{x/\sqrt{6}}
+ {\hat f}_0^{(2)}~{\rm e}^{2x/\sqrt{6}} + \ldots
\end{equation} 
and similarly for ${\hat q}_0(x)$ and ${\hat \mu}_0(x)$.   
 Eqs.~(\ref{order-u}) can be satisfied if the asymptotic form
of ${\hat f}_1(x)$ is 
\begin{eqnarray}
{\hat f}_1(x) = {\hat f}_1^{(0)}+&&\left({\hat f}_1^{(1)}+
{\hat g}_1^{(1)} \hat x \right)~{\rm e}^{x/\sqrt{6}} \nonumber \\
&&+ \left({\hat f}_1^{(2)}+{\hat g}_1^{(2)} \hat x \right)
~{\rm e}^{2x/\sqrt{6}} + \ldots, 
\end{eqnarray}
and similarly for ${\hat q}_1(x)$ and ${\hat \mu}_1(x)$. 
 At $O(u^2)$, ${\hat f}_2(x)$ would have second-order polynomials
multiplying the exponentials, and so on. 
 Substituting these expressions into the differential equations
allows us to determine the unknown constants (except for those
associated with the translational invariance). 
 For $f_{\infty}$ it yields the series 
\begin{equation}
f_{\infty}= \sqrt{\frac{2}{3}} + \frac{u}{24 \sqrt{6}} +
+ \frac{u^2}{768 \sqrt{6}} + \ldots, 
\end{equation}
which corresponds to
\begin{equation}
J^*=\frac{2}{3 \sqrt{3}} -\frac{u^2}{ 576 \sqrt{3}} 
-\frac{u^3}{5184 \sqrt{3}} + \ldots. 
\label{small_u_series}
\end{equation}
Note that the first correction to the $u \rightarrow 0$ limit
of $J^*$ is of $O(u^2)$, since the lowest term $J_c$
is at the maximum of $J^*(f_{\infty})=f_{\infty}^2
\sqrt{1-f_{\infty}^2}$. 
 
\begin{figure}
\centerline{
\epsfxsize=8.5cm
\leavevmode \epsfbox{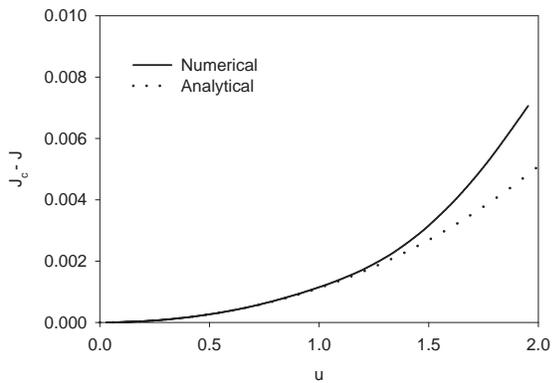}}
\caption{$(J_c-J)$ for the numerical data (solid 
line) and for the result of the small-$u$ analysis,
Eq.~(\ref{small_u_stall}) (dashed line).} 
\label{small-u-fig}
\end{figure}

The series found through the asymptotic perturbative expansion
above can be obtained by another method.
 Looking back at Eqs.~(\ref{sol1}) and (\ref{sol2}), we note that
the ratio of decay lengths $\lambda_f/\lambda_{\mu}=2$. 
 If we insert the expressions we have for these length scales,
Eqs.~(\ref{approach-3})
and (\ref{LA-length}), we find as $u \rightarrow 0$  
\begin{equation}
\frac{ \lambda_{f}}{\lambda_{\mu}}= 
\left[ \frac{uf_{\infty}^2}{6f_{\infty}^2-4}\right]^{1/2}=2. 
\end{equation} 
Solving for $f_{\infty}$, and then calculating $J^*$, we find 
\begin{equation}
J^*= J_c (1-u/8)^{1/2} (1-u/24)^{-3/2}, 
\label{small_u_stall}
\end{equation}
with $J_c= \sqrt{4/27}$, which when expanded for small-$u$ agrees
with the series (\ref{small_u_series}) found above. 
 We plot the small-$u$ numerical data and this expression together
in Fig.~\ref{small-u-fig}.
 The fit is surprisingly good at small $u$, suggesting to us that 
the corrections to Eq.~(\ref{small_u_stall}) are exponentially small
as $u\rightarrow 0$. 

\section{Moving Interfaces} 
\label{moving} 

At currents other than $J^*$, the NS interfaces move with a constant
velocity. 
 For such solutions the operator $\partial_t$ can be replaced by
$-c\partial_x$, so that Eqs.~(\ref{analytic}) become
\begin{mathletters}
\label{speed}
\begin{eqnarray}
&& -cuf_x = f_{xx} - f q^2  +f - f^3 , 
\label{speed-1} \\
&&u(-cq+\mu )f= 2f_x q + f q_x, 
\label{speed-2} \\
&& J =  f^2 q - \mu_x. 
\label{speed-3}
\end{eqnarray}
\end{mathletters}   
While the boundary conditions on $f$ and $q$ remain the same, 
that on the scalar potential becomes $\mu_{\infty}=cq_{\infty}$.  
 Actually, it is more convenient to use instead $\tilde \mu =
\mu -c q$, which is the gauge-invariant potential in the 
constant-velocity case.  

\begin{figure}
\centerline{
\epsfxsize=8.5cm
\leavevmode \epsfbox{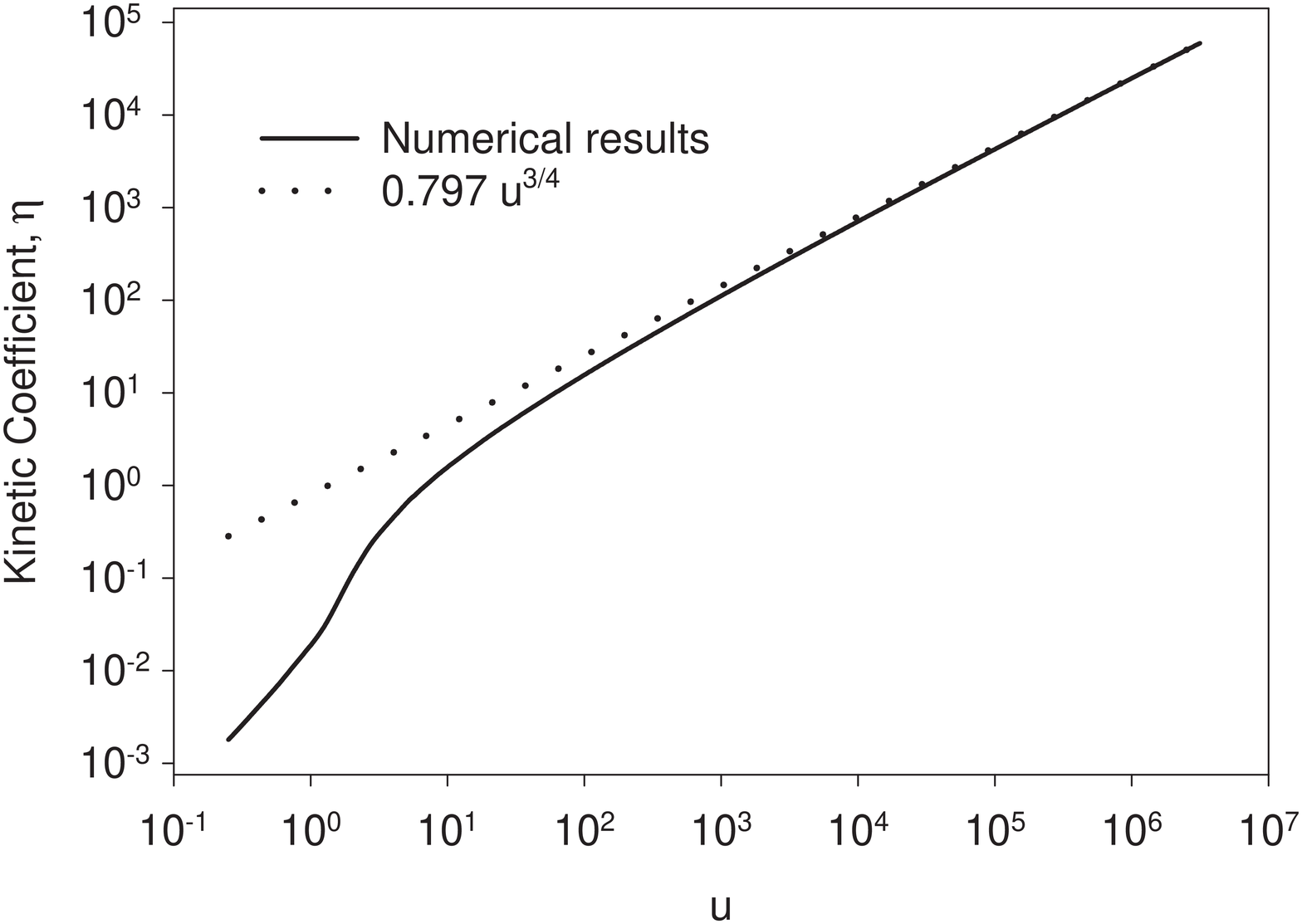}}
\caption{A log-log plot of the numerically determined kinetic
coefficient as a function of $u$ (the solid line) along with
an asymptotic fit of $0.797 u^{3/4}$ (the dotted line).  }
\label{fig5}
\end{figure}

The superconducting phase invades the normal phase if $J < J^*$ and
vice versa if $J > J^*$. 
 For currents near $J^*$, the interface speed is proportional to
$(J-J^*)$. 
 In this linear response regime, one can define a kinetic coefficient
(which Likharev \cite{likharev74} refers to as a ``viscosity") 
\begin{equation}
\eta = \left( {dc \over dJ } \right)^{-1}_{J=J^*}. 
\end{equation}  
Figure~\ref{fig5} shows the numerically determined kinetic coefficient 
as a function of $u$. 
 For large-$u$, we find $\eta \sim u^{3/4}$ for which we provide an 
argument below.   


\begin{figure}
\centerline{
\epsfxsize=8.5cm
\leavevmode \epsfbox{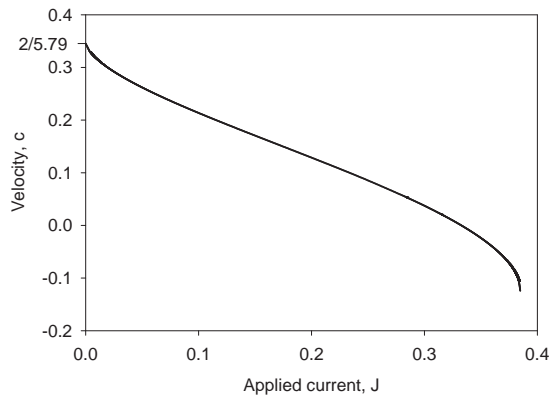}}
\caption{The velocity of the front versus the current $J$ 
for $u=5.79$.  }
\label{fig6}
\end{figure}

Farther from  the stall current, the velocities deviate from this
linear behavior, as seen in Fig.~\ref{fig6}. 
 The greatest departure occurs in the limits $J \rightarrow 0$
and $ J \rightarrow J_c$. 
 In fact, Likharev \cite{likharev74} conjectured that the interface
speed diverges in both of these limits; we find that it is bounded.    

{\it The $J \rightarrow 0$ limit}. 
The moving interface equations, Eqs.~(\ref{speed}), simplify in
the $J \rightarrow 0$ limit, since that limit implies that
both $q \rightarrow 0$ and $\mu \rightarrow 0$, leaving only 
\begin{equation}
f_{xx} + uc f_x +f -f^3 = 0. 
\label{fisher-kpp} 
\end{equation}
If we replace $uc$ in the above equation by a speed $v$, then
we have the steady-state version of Fisher's equation
\cite{fisher-kpp}, which is known to have propagating front
solutions with $v=2$ \cite{aronson78}. 
 In our case this implies that as $J \rightarrow 0$, $c=2/u$,
which is in good agreement with the numerical data shown in
Fig.~\ref{fig6}.     

We can combine the above result with an earlier one to suggest
that $\eta\sim u^{3/4}$ as $u \rightarrow \infty$. 
 In the large-$u$ limit, we have information on the following two
points: (1)  the stalled interface ($J=J^*\sim u^{-1/4},c=0$); and
(2) the interface in the absence of current ($J=0,c=2/u$). 
 In going from (1) to (2), the changes in current and velocity are
$\Delta J \sim u^{-1/4}$ and $\Delta c  \sim u^{-1}$. 
 As $u \rightarrow \infty$, both of these changes are small so that 
$\eta$ might be approximated by 
\begin{equation}
\eta \approx {\Delta J \over \Delta c } \sim  u^{3/4},  
\end{equation}
yielding the behavior seen in the numerical data
(see Fig.~\ref{fig5} and Table~\ref{table2}).

\begin{figure}
\centerline{
\epsfxsize=8.5cm
\leavevmode \epsfbox{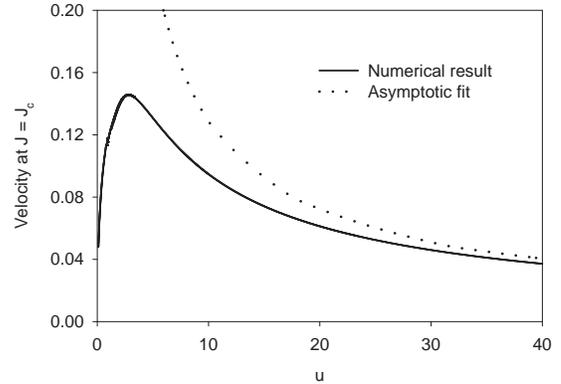}}
\caption{The velocity as $J \rightarrow J_c$ as a function of 
$u$.  For large $u$, the velocity asymptotically approaches
$0.92 u^{-0.85}$.}
\label{fig7}
\end{figure}

{\it The $J \rightarrow J_c$ limit.}
 The numerical work indicates that the velocity is finite as
$J \rightarrow J_c$; the limiting velocity is shown in
Fig.~\ref{fig7} as a function $u$.    
 We can find an analytic bound on this velocity as follows. 
 First, take Eqs.~(\ref{speed}), use the gauge-invariant potential
$\tilde \mu$, and find the constant-velocity analog of 
Eq.~(\ref{fandmu}). 
 Then substitute the asymptotic forms, Eqs.~(\ref{series}), into
the resulting equations, leading to   
\begin{mathletters}
\begin{eqnarray}
&&\left(cu\lambda_f^{-1} + \lambda_f^{-2}-2f_{\infty}^2 \right)
f_1 {\rm e}^{x/\lambda_f} = 2 f_{\infty}q_{\infty}q_1
{\rm e}^{x/\lambda_q}, \\
&&\left( uf_{\infty}^2 -\lambda_{\mu}^{-2}\right)\tilde \mu_1
{\rm e}^{x/\lambda_{\mu}} = c q_1 \lambda_q^{-2}
{\rm e}^{x/\lambda_q}, \\
&& 2f_{\infty}q_{\infty}f_1 {\rm e}^{x/\lambda_f} +
\left( f_{\infty}^2 -c \lambda_q^{-1}\right)q_1
{\rm e}^{x/\lambda_q} \nonumber \\
&& \ \ \ \ \ \ \qquad \qquad \qquad  
- \tilde \mu_1 \lambda_{\mu}^{-1}
{\rm e}^{x/\lambda_{\mu}}=0.
\end{eqnarray}
\end{mathletters}
Arguments similar to those following Eqs.~(\ref{approach}) lead one 
to the conclusion that in this case 
$\lambda_f=\lambda_q=\lambda_{\mu}$. 
 The above equations can then be shown to yield the following
relation 
\begin{eqnarray}
\label{c-bound}
&&u^2 c^2 +\left(2u\lambda^{-1}
-2uf_{\infty}^2\lambda -u^2 f_{\infty}^2 \lambda \right)c
\nonumber \\
&&+\left[2 \left(uf_{\infty}^2 \lambda^2-1 \right)
\left(3 f_{\infty}^2-2 \right)  -
uf_{\infty}^2 +\lambda^{-2}\right]=0,  
\end{eqnarray} 
where we have used $q_{\infty}^2=1-f_{\infty}^2$. 
 We find the bound by:  (1) solving Eq.~(\ref{c-bound}) for $c$;
(2) substituting in $f_{\infty}=\sqrt{2/3}$ (which corresponds to
$J=J_c$); and (3) extremizing that result with respect to the
decay length $\lambda$. 
 The small-$u$ limit of the resulting bound is $-\sqrt{2u}/9$,
and the large-$u$ limit is $-1/2\sqrt{3}$. 
 The square-root dependence of the velocity in the small-$u$
limit agrees with the data.  
 Now we can consider going from the stall current $(J^*,c=0)$
to the critical depairing current $(J_c,c\sim u^{1/2})$ which
results in changes $\Delta J \sim u^2$ and $\Delta c \sim u^{1/2}$,
suggesting that the small-$u$ kinetic coefficient $\eta \sim u^{3/2}$,
which is in rough agreement with the numerical data.        
 We have also observed that as a function of $J$ the speed appears
to approach its bound via a square root dependence         
\begin{equation}
c(J)= A + B (J_c-J)^{1/2}
\end{equation}
for all $u$. 


\section{Summary and Remarks}
\label{conclusions}

In this paper we have studied in detail the nucleation and 
growth of the superconducting phase in the presence of a current.
 The finite amplitude critical nuclei grow as the current is
increased, with the amplitude eventually saturating as the stall
current $J^*$  is approached, leading to the formation of interfaces
separating the normal and superconducting phases. 
 The stall current can be calculated in the limit of large $u$ using
matched asymptotic expansions, demonstrating once again the utility
of this technique for problems in inhomogeneous superconductivity.
 We have also derived an analytic expression for the stall current
for small $u$, which we believe to be correct up to exponentially
small corrections.
 Deviations from the stall current cause the interfaces to move, and
we have calculated the mobility of these moving interfaces for a wide
range of $u$.  
 Finally we have shown that the interface velocity $c= 2/u$ as
$J\rightarrow 0$ and that $c$ is bounded as $J\rightarrow J_c$, in
contrast to some conjectures in the literature.  

As in the magnetic-field analogy, the issue of stability and dynamics
of the current-induced NS interfaces will be more complicated and 
interesting in the two-dimensional case.
 Some preliminary work in this direction has been reported by
Aranson {\it et al.}\cite{aranson96}, who find that the current has a
stabilizing effect on the NS interface.
 This can be interpreted as a positive surface tension for the
interface, due entirely to {\it nonequilibrium} effects.
 They provide a heuristic derivation of an interesting free-boundary
problem for the interfacial dynamics (a variant of the Laplacian
growth problem);  however, this free-boundary problem is sufficiently
complicated that they were unable to solve it to compare with their
numerical results. 
 Clearly, further work in this direction would be helpful in 
understanding the nucleation and growth of the superconducting 
phase in two-dimensional superconducting films.

\acknowledgments
This work was supported by NSF Grants DMR 96-28926 and DMR 93-12476.

\appendix
\section{Amplitude of the critical nuclei in the 
$J \rightarrow 0$ limit}
\label{appendix}

In this appendix we provide a self-consistent calculation of
the amplitude of the critical nuclei in the $J \rightarrow 0$
limit. 
 Choosing the gauge appropriate for bumps centered at $x=0$
and combining Eqs.~(\ref{scaled}) into one equation yields 
\begin{eqnarray}
\label{start}
&&\left[-u \partial_t +iuJx + \partial_x^2  +1 \right] \psi(x,t)
= |\psi(x,t)|^2~ \psi(x,t) 
\nonumber \\
&& \ \ \ \ \ \ 
+iu \left[\int_0^{x} dy ~{\rm Im} \left(\psi^*(y,t)
\partial_y \psi (y,t) \right) \right] \psi(x,t).
\end{eqnarray} 
The propagator for the linear operator appearing on the left hand
side of Eq.~(\ref{start}) satisfies the condition   
\begin{eqnarray}
\left[-u \partial_t  +iuJx + \partial_x^2  +1
\right]&& G(x,x';t-t') \Theta(t-t') 
\nonumber \\ 
= &&-u~ \delta(x-x') \delta(t-t')  
\end{eqnarray} 
and is given by  
\begin{eqnarray}
G(x,x';\tau) = &&\left( \frac{u}{4 \pi \tau} \right)^{1/2} 
\exp \left[ \frac{\tau}{u} - \frac{J^2 \tau^3}{12u} \right]
\nonumber \\
&& \times \exp \left[ \frac{iJ \tau(x+x')}{2} -
\frac{u(x-x')^2}{4 \tau} \right] . 
\end{eqnarray} 
Ivlev {\it et al.} \cite{ivlev80,ivlev84} used this linear
propagator to evolve perturbations having widths of $O(1)$ and
carrying no current. 
 Without the nonlinear terms such perturbations initially grow
but ultimately reach a maximum size and then decay away.
 Ivlev {\it et al.} suggested that the amplitudes of the critical
nuclei are exponentially small in the $J \rightarrow 0$ limit by
asking what sized initial perturbations are of $O(1)$ at their
maxima.  
 Their arguments motivated us to use the propagator in a more
careful estimate of the amplitude that includes the nonlinear
terms as an essential ingredient. 

We can convert Eq.~(\ref{start}) into an integral equation by 
multiplying both sides of Eq.~(\ref{start}) (with $x \rightarrow
x'$ and $t \rightarrow t'$) by $ G(x,x';t-t')$ and integrating
over all $x'$ and integrating $t'$ from $0$ to $t$. 
 After some manipulation these steps lead to  
\begin{eqnarray}
\label{final}
 \psi(x,t) && =  
\int_0^t dt' \int_{-\infty}^{\infty} dx'  ~G(x,x',t-t') 
\nonumber \\
\times&&\left\{
\psi(x',t')\delta(t-t')  -\frac{1}{u}  
|\psi(x',t')|^2~ \psi(x',t') \right. \nonumber \\ 
 -i  && \left. \left[\int_0^{x'} dy ~{\rm Im} \left( 
\psi^*(y,t') \partial_y \psi(y,t')
\right) \right] \psi(x',t') \right\} , 
\end{eqnarray}
where $t>0$.

In order to estimate the amplitude of the threshold solutions, we
will substitute into Eq.~(\ref{final}) the following form
\begin{equation}
\label{form}
\psi(x,t) = \psi_0 \exp \left\{  -\frac{uJx^2}{4} + ix 
\right\}.
\end{equation} 
Note that this form is stationary and has a fixed Gaussian shape 
(which is inspired by our WKB approximation, see Eq.~(\ref{4})) 
but it has an arbitrary amplitude which we will determine
self-consistently. 

Let us take the $t \rightarrow \infty$ limit and focus on $x=0$
since our interest is in the amplitude. 
 After substituting Eq.~(\ref{form}) into Eq.~(\ref{final}), we
can do both integrals for the first term on the right hand side
exactly, and it can be seen to decay to zero in the $t \rightarrow
\infty$ limit.  
 Next, we perform the $x'$ integration of the second term on the
right hand side ($II$), which yields 
\begin{equation}
II=-\frac{\psi_0^3}{uJ} \int_0^{\infty} \frac{d\tau}{\sqrt{1+3\tau}}
\exp \left\{ \frac{24\tau^2-4\tau^3-3\tau^4}{12uJ(1+3\tau)} \right\},
\end{equation}
where $\tau=Jt$. 
 We now apply the method of steepest descent to obtain 
\begin{equation}
II \approx   -\frac{\sqrt{\pi}\psi_0^3}{\sqrt{2uJ}} \exp \left\{
\frac{32}{81~uJ} \right\} .
\end{equation}

In the third term on the right hand side ($III$) of Eq.~(\ref{final}),
we make the substitution $y=vx'$ and then perform the $x'$ integration 
giving 
\begin{eqnarray}
III & = & \frac{\psi_0^3}{uJ^2}\int_0^{\infty} d\tau  \int_0^1 dv  
\frac{(2\tau+\tau^2)}{\sqrt{[1+\tau(1+2v^2)]^3}} \nonumber \\
&& \times \exp 
\left\{ \frac{24v^2\tau^2-4\tau^3-(1+2v^2)\tau^4}
{12uJ[1+(1+2v^2)\tau]}\right\}. 
\end{eqnarray}
The maximum of the term in the exponential of $III$ occurs at 
$v=1$ (which is an endpoint). 
 Linearizing about that maximum provides  
\begin{eqnarray}
III & \approx & \frac{\psi_0^3}{uJ^2} \int_0^{\infty} 
\frac{ d \tau \tau(2+\tau)}
{\sqrt{(1+3 \tau)^3}} \exp 
\left\{ \frac{24\tau^2-4\tau^3-3\tau^4}{12uJ(1+3\tau)}
\right\} 
\nonumber \\
&& \times \int_0^1 dw \exp \left\{ -\frac{\tau^2(2+\tau)^2w}
{uJ(1+3\tau)^2}\right\},
\end{eqnarray}
where $w=1-v$. 
 After the $w$ integration, we apply the method of steepest 
descent to the $\tau$ integration to obtain   
\begin{equation}
III \approx \frac{\sqrt{\pi u}\psi_0^3}{\sqrt{2J}} 
\frac{9}{8} \exp 
\left\{ \frac{32}{81~uJ} \right\}.
\end{equation}
Putting all of these results back into Eq.~(\ref{final}) gives 
\begin{equation}
\psi_0 \approx  
\frac{\sqrt{\pi u}\psi_0^3}{\sqrt{2J}} 
\exp \left\{ \frac{32}{81~uJ} \right\}
\left[\frac{9}{8}  - \frac{1}{u} \right],  
\end{equation} 
which provides the expression given in the text, Eq.~(\ref{self}). 
 This calculation clearly runs into trouble when $u<8/9$; however,
the numerical coefficients in front of these integrals are 
sub-leading terms, and they can be varied by adding sub-leading 
terms to the initial Gaussian guess.




\begin{references}

\bibitem[*]{emaild} Electronic address: ajd2m@virginia.edu

\bibitem[*]{emailt} Electronic address: teb2n@virginia.edu

\bibitem[*]{emaila} Electronic address: dorsey@phys.ufl.edu 

\bibitem[*]{emailm} Electronic address: mf1i@virginia.edu

\bibitem{frahm91}H.\ Frahm, S.\ Ullah, and A.\ T.\ Dorsey, Phys.\ 
Rev.\ Lett.\ {\bf 66}, 3067 (1991). 

\bibitem{liu91}F.\ Liu, M.\ Mondello, and N.\ D.\ Goldenfeld, Phys.\ 
Rev.\ Lett.\ {\bf 66}, 3071 (1991). 

\bibitem{dibartolo96}S.\ J.\ Di~Bartolo and A.\ T.\ Dorsey, Phys.\ Rev.\ 
Lett.\ {\bf 77}, 4442 (1996). 

\bibitem{dorsey94}A.\ T.\ Dorsey, Ann.\ Phys.\ {\bf 233},
248 (1994).

\bibitem{osborn94}J.\ C.\ Osborn and A.\ T.\ Dorsey, Phys.\ Rev.\ B 
{\bf 50}, 15~961 (1994).  See also  C.\ J.\ Boulter and J.\ O.\ Indekeu, 
Phys.\ Rev.\ B {\bf 54}, 12~407 (1996). 

\bibitem{chapman95a}S.\ J.\ Chapman, Quart.\ J.\ Appl.\ Math.\
{\bf 53}, 601 (1995).

\bibitem{pippard50}A.\ B.\ Pippard, Phil.\ Mag.\ {\bf 41}, 243 (1950).

\bibitem{chapman95b}S.\ J.\ Chapman, IMA J.\ Appl.\ Math.\ {\bf 54}, 
159 (1995). 

\bibitem{goldstein96}R.\ E.\ Goldstein, D.\ P.\ Jackson, and A.\ T.\ Dorsey,
Phys.\ Rev.\ Lett.\ {\bf 76}, 3818 (1996); A.\ T.\ Dorsey and R.\ E.\ 
Goldstein, preprint (cond-mat/9704161). 

\bibitem{likharev74}K.\ K.\ Likharev, JETP Lett.\ {\bf 20},
338 (1974).

\bibitem{ivlev84}B.\ I.\ Ivlev and N.\ B.\ Kopnin, Sov.\ Phys.\ Usp.\ 
{\bf 27}, 206 (1984).

\bibitem{kramer77}L.\ Kramer and A.\ Baratoff, Phys.\ Rev.\ Lett.\
{\bf 38}, 518 (1977).

\bibitem{aranson96}I.\ Aranson, B.\ Ya.\ Shapiro, and V.\ Vinokur, Phys.\ 
Rev.\ Lett.\ {\bf 76}, 142 (1996). 

\bibitem{schmid66}A.\ Schmid, Phys.\ Kondens.\ Mater.\ {\bf 5}, 
302 (1966). 

\bibitem{gorkov68}L.\ P.\ Gor'kov and G.\ M.\ Eliashberg, Sov.\ Phys.\ 
JETP {\bf 27}, 328 (1968). 

\bibitem{mae}For a general introduction to the method of matched 
asymptotic expansions, see  
C.\ M.\ Bender and S.\ A.\ Orszag, {\it Advanced Mathematical
Methods for Scientists and Engineers} (McGraw-Hill, New York 1978);
M. van Dyke, {\it Perturbation Methods in Fluid
Mechanics} (The Parabolic Press, Stanford 1975). For some recent
applications to problems in superconductivity, see 
Refs.~\cite{dorsey94} and \cite{chapman95a}, and A.\ J.\ Dolgert,
S.\ J.\ 
Di~Bartolo, and A.\ T.\ Dorsey, Phys.\ Rev.\ B {\bf 53}, 5650 (1996). 

\bibitem{pearl}J. Pearl, Appl.\ Phys.\ Lett.\ {\bf 5}, 65 (1964).  

\bibitem{chapman97}S.\ J.\ Chapman, Q.\ Du, and M.\ Gunzburger, Z.\ 
angew.\ Math.\ Phys.\ {\bf 47}, 410 (1997). 

\bibitem{length}We warn the reader that this choice is different than 
in many applications of the TDGL equations, where the 
penetration depth is chosen as the length scale.  In the thin film
limit the penetration depth drops out of the calculation, leaving
$\xi$ as the length scale. 

\bibitem{ivlev80b}B.\ I.\ Ivlev, N.\ B.\ Kopnin and L.\ A.\ Maslova, 
Sov.\ Phys.\ Solid State {\bf 22}, 149 (1980). 

\bibitem{gorkov70}L.\ P.\ Gor'kov, JETP Lett.\ {\bf 11}, 32 (1972). 

\bibitem{kulik70}I.\ O.\ Kulik, Sov.\ Phys.\ JETP {\bf 32}, 318 (1970). 

\bibitem{num-rec}W.\ H.\ Press, B.\ P.\ Flannery, S.\ A.\ Teukolsky and
W.\ T.\ Vetterling, {\it Numerical Recipes} (Cambridge Univ. Press,
New York 1986).

\bibitem{watts-tobin81}R.\ J.\ Watts-Tobin, Y.\ Kr\"ahenb\"uhl 
and L.\ Kramer, J.\ Low Temp.\ Phys. {\bf 42} 459 (1981). 

\bibitem{ivlev80}B.\ I.\ Ivlev, N.\ B.\ Kopnin and L.\ A.\ Maslova, 
Sov.\ Phys.\ JETP {\bf 51}, 986 (1980).  



\bibitem{langer67}J.\ S.\ Langer and V.\ Ambegaokar, Phys.\ Rev.\
{\bf 164}, 498 (1967).

\bibitem{fisher-kpp}R.\ A.\ Fisher, Ann.\ Eugen.\ {\bf 7}, 355 (1937);
A.\ N.\ Kolmogorov, I.\ G.\ Petrovskii and N.\ S.\ Pishunov, Bull. Univ.
Moscow, Ser. Int. A {\bf 1}, 1 (1937).

\bibitem{aronson78}D.\ G.\ Aronson and H.\ F.\ Weinberger, Adv.\ Math.\
{\bf 30}, 33 (1978).

\end{references}
\end{document}